 \documentclass[]{spieScot}  

 \usepackage{graphicx}
 \usepackage{amssymb}
 \usepackage{epstopdf}
 \usepackage{times}	
 
\def\note#1{\relax} 

\title{TEDI: the TripleSpec Exoplanet Discovery Instrument} 

\author{Jerry Edelstein\supit{a}, Matthew Ward Muterspaugh\supit{a},
David J. Erskine\supit{b}, \\
 W. Michael Feuerstein\supit{a}, Mario Marckwordt\supit{a}, Ed Wishnow\supit{a}, \\
James P. Lloyd\supit{c}, Terry Herter\supit{c}, Phillip Muirhead\supit{c}, George E. Gull\supit{c}, \\
Charles Henderson\supit{c}, Stephen C. Parshley\supit{c} 
\skiplinehalf
\supit{a}Space Sciences Lab. at Univ. of Calif., Berkeley, CA 94720-7450\\
\supit{b}Lawrence Livermore Nat. Lab., 7000 East Ave, Livermore, CA 94550 \\
\supit{c}Astronomy Dept., Cornell University, Ithaca, NY 14853 }

\authorinfo{ jerrye@ssl.berkeley.edu, 510-642-0599}

\begin{document}
 \maketitle
 
{\bf Copyright 2007 Society of Photo-Optical Instrumentation Engineers.\\}
This paper will be published in SPIE conference proceedings volume 6693, 
``Techniques and Instrumentation for Detection of Exoplanets III.''  
and is made available as 
and electronic preprint with permission of SPIE.  One print or electronic 
copy may be made for personal use only.  Systematic or multiple reproduction, 
distribution to multiple locations via electronic or other means, duplication 
of any material in this paper for a fee or for commercial purposes, or 
modification of the content of the paper are prohibited.

\begin{abstract}
The TEDI (TripleSpec--Exoplanet Discovery Instrument) will be the first instrument fielded specifically for finding low-mass stellar companions. The instrument is a near infra-red interferometric spectrometer used as a radial velocimeter. TEDI joins Externally Dispersed Interferometery (EDI) with an efficient, medium-resolution, near IR (0.9 -  2.4 micron) echelle spectrometer, TripleSpec,  at the Palomar 200'' telescope. We describe the instrument and its radial velocimetry demonstration program to observe cool stars.

\end{abstract}



\section{INTRODUCTION}

The TEDI (TripleSpec ╨ Exoplanet Discovery Instrument)
will be the first instrument fielded specifically for the
near infra-red (NIR) radial velocity (RV or Doppler) search for planetary companions
about low-mass stars.  These stars are bright in the NIR
and rich with the absorption features needed for precise RV measurements.
The primary and standard method for the detection of extrasolar planets
is the Doppler technique. 
Doppler methods measure the stellar radial velocity change imposed
by the reflex motion of an orbiting planet -- the RV change causes 
spectral shifts of stellar absorption line centroids. 
RV planet searches have, so far, been limited almost entirely to the optical band and
to solar-type stars that are bright in this band.
Consequently, knowledge about planetary companions to the populous but visibly-faint low mass stars  is limited.
Studying exo-planets about low mass stars is important because their distribution
discriminates among planet formation theories,
and because low-mass stars can more readily reveal very low-mass planets through measurable Doppler shifts.

The measurement of planetary induced RV effects is extremely challenging.
Typical planetary RV shifts are 1000 times smaller than the stellar spectral line widths
and
require extremely stable and well-calibrated instruments
 in order to separate the signal from systematic errors or drifts.
 Conventional RV measurements typically
 use high-resolution echelle spectrographs (R~50-100k)
 to obtain the required velocity precision.
These costly instrument use large optics and multi-meter paths that must be held
 to extreme tolerances, and maintaining their long-term stability often requires 
 substantial effort such as complex servo systems or an enclosing vacuum environment 
 \cite{Mayor1995,Vogt1987,Vogt1994}. 
 
TEDI uses Externally Dispersed Interferometry (EDI),  a combination of interferometry 
and multichannel dispersive spectroscopy,
that greatly improves the velocity resolution of moderate resolution, high throughput spectrographs. 
The TEDI instrument joins EDI with the Cornell TripleSpec infrared 
simultaneous (0.9 -  2.4 micron)-band spectrograph \cite{Wilson2004} at the Palomar Observatory 200╙ telescope.
We intend to conduct a science-demonstration program 
for the Doppler-search of planets orbiting low mass faint M, L and T type stars and brown dwarfs.  
 
 \section{Externally Dispersed Interferometry}

The externally dispersed interferometer 
\cite{ErskineGe2000,G.E.R.2002,ErskineEDITheory2003,ErskinePatSuperimpose,ErskinePatEDI2002,Ge2002,Ge2003,ResBoostApJ2003}
increases the Doppler sensitivity and multiplies the spectral resolution of an existing spectrograph over its full and simultaneous bandwidth by a factor of several to an order of magnitude (Erskine et al., 2003). For example, our prototype observatory and laboratory instruments have demonstrated EDI visible-band velocimetry precision to ~5 m/s using an R= 20,000 spectrograph, and a factor of six increase in conventional spectrograph resolving power (from R= 25k to R= 140k)\cite{SPIEscot}.

The EDI uses a series combination of a small fixed delay interferometer with a conventional grating spectrograph. 
The interferometer creates a transmission comb that is periodic with wavelength 
that multiplies the input spectrum to create moir\'e fringes.
These fringes provide a periodic spectral fiducial comb covering the entire bandwidth of the spectrograph. 
This comb is analogous to the fiducial lines of an iodine absorption cell, but with lines of exceedingly uniform spacing, shape, and amplitude over the entire bandwidth. The comb, in multiplication with the input spectrum, heterodynes fine spectral features into a lower spatial-frequency moir\'e pattern that is recorded by the spectrograph detector. 
The heterodyning can be numerically reversed to recover detailed spectral information otherwise unattainable by the spectrograph alone. 
The EDI fringing signal provides a precise internal fiducial that can be used to defeat systematic instrumental noise so that the tolerance to blur or pupil changes (Erskine \& Edelstein 2003)
is improved by orders of magnitude compared to
classical high-resolution spectrographs that directly map telescope image quality 
and pupil stability to spectral performance.

A Doppler velocity change induces a phase change in the moir\'e pattern relative to the moir\'e pattern of a simultaneously measured calibrant spectrum. 
In EDI, the moir\'e fringe phase becomes the primary diagnostic instead of classical spectral dispersion. 
 A vector data analysis procedure (Erskine 2003) precisely measures the differential moir\'e pattern phase between the input spectrum and a reference  spectrum (absorption cell or emission lamp) that has been simultaneously recorded in the same fringing signal. 
The moir\'e pattern has much broader features than the narrow stellar absorption lines that created it.  Hence a much lower resolution spectrograph can be used to make precision Doppler velocities than otherwise practical without the interferometer. In fact the native spectrograph can have a low resolution (3k to 5k) such that the stellar lines are not fully resolved.  Such an EDI has recently been used to discover an exoplanet in Virgo\cite{GeAAS2005}.

\section{The EDI-TripleSpec Instrument}

\begin{figure}[h]
\center{\includegraphics[height=10 cm]{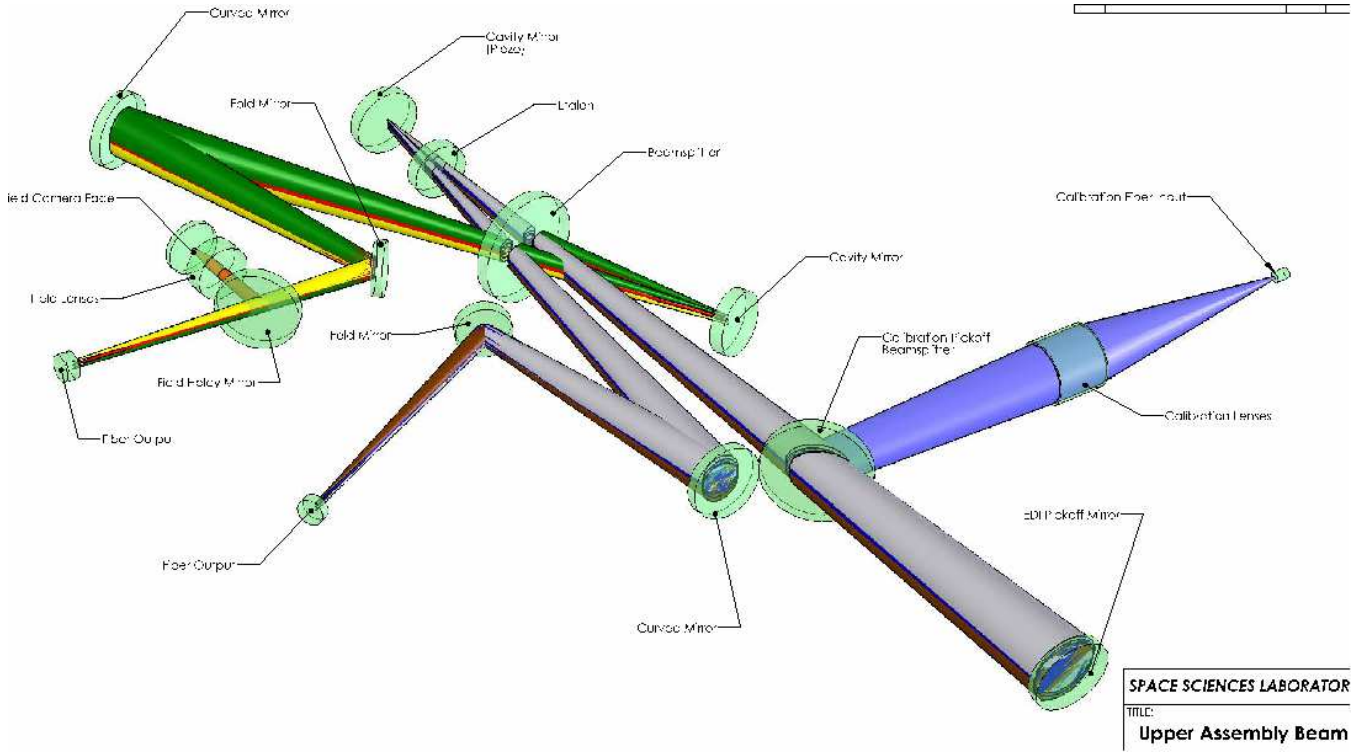}}
\caption[Fig1] {
{\bf TEDI Interferometer Optical Schema:}
External sources(star \& background) from the telescope 
enter from the lower right.
The beam transverses a 95/5\% beamsplitter where a calibration beam is injected
by way of a calibration source (upper right) and telescope.
The light goes through an unequal-arm off-axis Michelson
interferometer cavity, via the beam-splitter and cavity mirrors.
An etalon compensates the unequal-arm path difference,
providing large angular independence.
Folding ellipses and mirrors direct the two complementary-arm
output beams to optical fiber injection.
An alignment camera is focussed on one cavity mirror via a pickoff that
passes the science beam.}
\end{figure}

The TripleSpec EDI system consists of an interferometer (see schematic Fig.1, and illustration Fig.2) 
 mounted  to the TripleSpec entrance plate (within a 30 inch diameter volume available to the telescope Cassegrain feed). The interferometer consists of a fixed main optical unit and pick-off arm that can be moved to intercept the incoming f/16 telescope beam, divert it to the interferometer, and return the light to the beam path for TripleSpec.
The Diverted light from the telescope enters an interferometer cavity and focusses on the cavity mirrors.
The interferometer uses use an off-axis Michelson scheme that allows ready access to both the arms╒ outputs.
A delay is introduced in one arm using a controlled slide stage.
A selectable etalon in the delay arm creates a nearly angle-independent optical path to allow for a large field of view.
The beam-splitter and etalon are made of IR grade fused silica.

Recording fringes over a long exposure requires a stable phase robust to thermal and mechanical drifts. If  the phase  wanders more than  $\lambda / 8$, then the net visibility  will be significantly reduced and decrease S/N proportionally. Our data analysis can handle irregularly spaced phase steps, so only large  phase wandering has impact on
the net velocity resolution.
 We use a commercial PZT-transducer mounted mirror to actively compensate for optical path fluctuations. The system uses a visible gas-laser that propagates through the interferometer elements to form fringe patterns on a small camera. Software analyzes the location of the fringes and the PZT is moved to compensate. The same piezo-system is conveniently used for phase  stepping over observations in order to defeat instrumental systematic noise. 

A diamond-turned ellipse and folding flat direct each of the interferometer's complementary outputs 
to a linear array of three
100-micron diameter optical fibers made of fluoride glass and prefaced with 
a bonded micro-lens for coupling efficiency and scrambling.
The central fiber collects light from the target star.  
One outboard fiber collects light from an adjacent sky region for simultaneous background measurement.
The other outboard fiber collects light for a simultaneous, near common-path  calibration source input.
A selectable mixture of a ThAr emission lamp and a Halogen-filament continuum lamp
is supplied as the calibration source via fiber injection to an achromatic telescope made with 
an air-spaced SF6-Barium fluoride doublet.
The calibration telescope injects light to interferometer via an uncoated fused-silica plate 
placed in the main-telescope's input beam.

The two sets of fibers from the complementary outputs are collected into a single 
linear array of 6 fibers gor injection into the spectrograph.
  Each fiber's output is scrambled and converted to a slow beam
using a microlens.  The parallel fiber outputs are collimated by an achromatic cemented
doublet (SF6 and Barium fluoride). The collimated beams pass through a 'rotisserie'
that allows for calibration absorption cell insertion.  The cells, held at 35 C and 
containing methane, ammonia, and hydrogen sulfide, 
use IR fused silica windows that are wedged to avoid etalon effects. 
After the rotisserie, the beams are re-focussed using the identical cemented achromat
and directed to the TripleSpec entrance slit. 
Controlled staging allows for alignment and focus of optical paths both within TEDI and from
TEDI to the spectrograph.
Beam splitter, anti-reflection, and mirror coatings were selected for optical performance over the working band.
 The optical result is a set of 6 image points,
each of which is dispersed in parallel within each of the spectrograph's orders.

\begin{figure}[h]
\center{\includegraphics[height=10 cm]{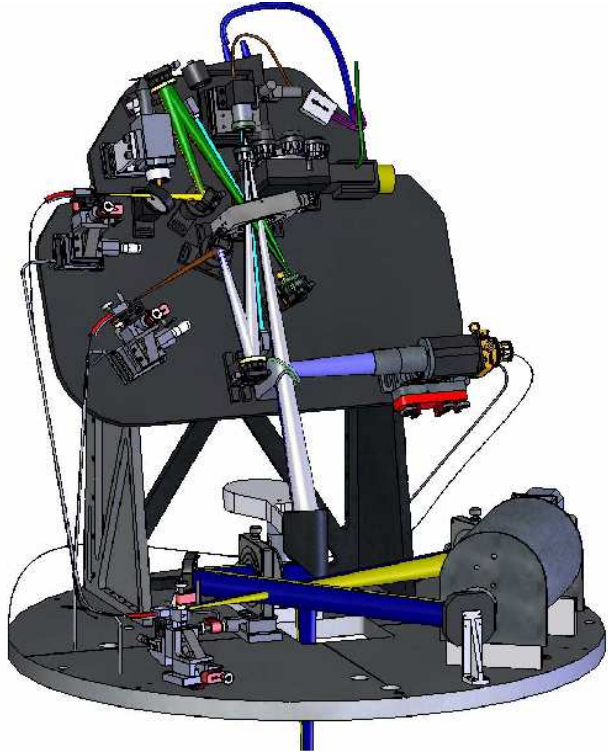} }
\caption[Fig2] {
{\bf The T-EDI Instrument}
The TEDI Palomar mechanical layout.
The telescope beam (not shown) enters from the top
along the base plate axis.
The upper plate holds the interferometer cavity
which consists of a beam splitter and two mirrors.
An selectable etalon is inserted before one of the mirrors.
A calibration source is injected into the input path.
A camera system stabilizes fringes formed by from a tracking laser 
traversing the cavity.
The complementary outputs are each folded an refocused into optical fibers.
The fibers are routed to the base plate
where they emit and are lens collimated to a beam 
in which an reference absorption cell can be inserted.
The beam is refocussed to the spectrograph entrance
which is centered below the base plate.
In this scaled perspective illustration, the base plate is 30 inches across.
}
\end{figure}

\section{Radial Velocimetry Reference Standards}

The introduction of the Iodine absorption-line reference cell was crucial to the current renaissance in precision radial velocity studies and to the detection of extrasolar planets. The reference cell provides a stable zero-velocity narrow-line calibrant needed to precisely determine slight Doppler shifts over long time scales. Reference cells are still central to radial velocity studies today, although although  some groups now use a Th-Ar emission lamp or a laster stabilized spectrographs instead. Conventional Doppler measurements are limited to bands where the standards have adequate transmission or dense and narrow spectral signatures.

The atmosphere itself provides a dense comb of NIR narrow absorption lines, similar to that of an absorption cell. 
Using the detailed atmospheric absorption models of Roe (2002) we find that the atmospheric spectrum 
could have adequate spectral feature density and slope for use as a velocimetry absorption cell at precision of $\sim$ 10 m/s. However, atmospheric variations, including line sight velocity perturbations, will constrain the usable velocity.
Atmospheric lines other than the water vapor lines are not expected to change dramatically. 
Because EDI determinations depends on velocity centroids of reference lines, and not the reference line profile
we may be able to use non-aqueous telluric lines for limited RV reference 
as their centroids are unlikely altered by optical depth effects. 

Rather than depend on the atmosphere, we intend to use a the rich ThAr near-IR
emission lamp spectrum
as the primary reference standard. Injection of the emission-lamp flux into an EDI is simplified
in comparison to methods used for conventional spectroscopy 
because our angle-independent interferometer design imprints the same spectral fiducial comb 
upon both the stellar and lamp beams even if they are not exactly co-axial,
and because of the insensitivity of EDI to pupil and point-spread function illumination.
Furthermore, EDI╒s transmission comb will form beat patterns with spectra 
to provide an absolute spectral reference frame that is repeatable throughout all wavelengths.

While many gases have  have useful NIR absorption signatures,
they do not make for efficient, simultaneous primary velocity standards because
their information content is so limited in comparison to that of the feature-rich stars we wish to measure.
Therefore we intend to employ gas absorption cells as a secondary transfer standard and to facilitate alignment.
For example, to calibrate differences between the optical paths used for the star, background,
and calibrations beams, we will record the simultaneous
response of white light passing through each optical chains
 and an absorption cell.
 
\section{Sensitivity}

We have estimated the sensitivity and velocity noise for TEDI.
Because the EDI process recovers both a conventional and a fringing spectrum,
both can be  combined to form a net EDI response. 
The two spectral components are determined from different spatial frequencies 
on the detector and are statistically independent with uncorrelated errors.
The net EDI response is then formed from the component's quadrature sum. 
The systematic errors suffered by the conventional and the fringing spectrum are
different however. The EDI derived velocity is orders of magnitude lower than the
 common limiting systematic errors for the conventional spectra,
such as variation in pupil illumination or point spread function.

We estimated the RV precision using TEDI for a 5 minute exposure of 
a modeled star (t = 2400 Kelvin, g = 0.5) with M$_{H}$ = 10,
with zero rotational velocity,
given a Gaussian R=2,700 spectrograph response,
 an interferometer delay best matched to the Doppler information content,
 and throughputs estimated for the TripleSpec spectrograph and the TEDI instrument
at the Palomar 200╙ telescope.  
The analysis includes considerations for noise and dynamic range,
and regions of strong telluric contamination were excluded
from the sensitivity analysis.
 Our results (see Tab. 1) show that the EDI RV precision for each spectrograph order is about 10 m/s
 and that the orders can be combined to a single velocity measurement with a precision of 3.2 m/s.
The net velocity noise is effected by the noise in determining both the telluric line components and the reference line components.
 High-resolution model atmosphere specta show that the telluric lines will not add significant velocity noise.
 Our calculations using measured ThAr spectra (shown in Table 1) show that the reference determination
adds 1 m/s to the net RV precision of 4 m/s.
Stellar rotational velocity limits Doppler precision for any measurment.
For example, with the same parameters as above but v = V*sin(i) = 6 km/s,
the net RV precision is estimated to be 7 m/s.
 
\begin{center}
 \begin{tabular}{|cccc|}
\hline
     Order &   Bandpass &     ThArRV &     StarRV \\

           &     micron &        m/s &        m/s \\
\hline
         1 & 0.8 - 1.06 &        0.6 &       10.4 \\
         2 & 0.94 - 1.25 &        0.5 &       10.3 \\
         3 & 1.13 - 1.48 &        0.5 &       10.1 \\
         4 & 1.41 - 1.86 &        0.9 &        9.4 \\
         5 & 1.87 - 2.4 &        1.5 &        9.9 \\
\hline
           &   Combined &        0.9 &        3.2 \\
\hline
           &        \bf{Net} &            &        \bf{4.1} \\
\hline
           \bf{Table I:} & \bf{Estimated} & \bf{Sensitivity} & \\
\hline

\end{tabular}
 \end{center}

 Our visible-light EDI instruments have already demonstrated a $<$ 5 m/s RV measurement noise performance in the observatory and in benchtop tests (Ge, Erskine \& Rushford 2002) over durations from 11 days to a month, and achieved near photon limited velocity noise, a remarkable result given the prototype instrument quality and absence of environmental controls. The proven duration of stability for EDI is more than sufficient for the 3 to 5 day periods that our observing plan is seeking to test. Nonetheless, we plan to conduct long term stability testing using a zero-velocity references
 in order to understand the systematic limitations of our method.

The TEDI is scheduled for deployment at Palomar at the end of 2007, with an
 demonstration program to follow commissioning.
The instrument science demonstration program will 
search for planets about nearby cool stars. Considering that 
100 m/s precision is sufficient to detect a Jupiter mass planet in a 3 day orbit 
about a star with with 0.3 Solar masses (typical of a mid M star),
then our estimated RV precision should suffice for the detection
of these and lighter planets.  We plan to observe 100 M dwarfs ( M$_{H} <$10),
25 L dwarfs (10 $< M_{H} <$  15), and 10 T dwarfs (13 $< M_{H} <$ 15) with 
an average of 5 samples per target, sampled at a cadence sufficient to observe 
few to 5-day periods.

\clearpage
\acknowledgments         
This material is based upon work supported by the National Science Foundation under Grant No. AST-0505366.


\bibliography{Mine2,OthersAstro2}   

\begin{thebibliography}{10}

\bibitem{Mayor1995}
M.~Mayor and D.~Queloz {\em Nature} {\bf 378}, p.~355, 1995.

\bibitem{Vogt1987}
S.~Vogt, ``{The Lick Observatory Hamilton Echelle Spectrometer},'' {\em PASP}
  {\bf 99}, p.~1214, 1987.

\bibitem{Vogt1994}
S.~Vogt {\em et~al.}, ``{HIRES: The High Resolution Echelle Spectrometer on the
  Keck Ten-Meter Telescope},'' {\em Proc. SPIE} {\bf 2198}, p.~362, 1994.

\bibitem{Wilson2004}
J.~C. {Wilson}, C.~P. {Henderson}, T.~L. {Herter}, K.~{Matthews}, M.~F.
  {Skrutskie}, J.~D. {Adams}, D.-S. {Moon}, R.~{Smith}, N.~{Gautier},
  M.~{Ressler}, B.~T. {Soifer}, S.~{Lin}, J.~{Howard}, J.~{LaMarr}, T.~M.
  {Stolberg}, and J.~{Zink}, ``{Mass producing an efficient NIR
  spectrograph},'' in {\em Ground-based Instrumentation for Astronomy. Edited
  by Alan F. M. Moorwood and Iye Masanori. Proceedings of the SPIE, Volume
  5492, pp. 1295-1305 (2004).},  A.~F.~M. {Moorwood} and M.~{Iye}, eds.,
  pp.~1295--1305, 2004.

\bibitem{ErskineGe2000}
D.~Erskine and J.~Ge, ``{Novel Interferometer Spectrometer for Sensitive
  Stellar Radial Velocimetry},'' in {\em {Imaging the Universe in Three
  Dimensions: Astrophysics with Advanced Multi-Wavelength Imaging Devices}},
  W.~van Breugel and J.~Bland-Hawthorn, eds., {\em ASP} {\bf 195}, p.~501,
  2000.

\bibitem{G.E.R.2002}
J.~Ge, D.~Erskine, and M.~Rushford, ``{An Externally Dispersed Interferometer
  for Sensitive Doppler Extra-solar Planet Searches},'' {\em PASP} {\bf 114},
  pp.~1016--1028, 2002.

\bibitem{ErskineEDITheory2003}
D.~Erskine, ``{An Externally Dispersed Interferometer Prototype for Sensitive
  Radial Velocimetry: Theory and Demonstration on Sunlight},'' {\em PASP} {\bf
  115}, pp.~255--269, 2003.

\bibitem{ErskinePatSuperimpose}
D.~Erskine, ``{Single and Double Superimposing Interferometer Systems},'' {\em
  US Patent} {\bf 6,115,121}, 2000.

\bibitem{ErskinePatEDI2002}
D.~Erskine, ``{Combined Dispersive/Interference Spectroscopy for Producing a
  Vector Spectrum},'' {\em US Patent} {\bf 6,351,307}, Feb. 26, 2002.

\bibitem{Ge2002}
J.~Ge, ``{Fixed Delay Interferometry for Doppler Extrasolar Planet
  Detection},'' {\em ApJ} {\bf 571}, pp.~L165--168, 2002.

\bibitem{Ge2003}
J.~Ge, ``{Erratum: Fixed Delay Interferometry for Doppler Extrasolar Planet
  Detection},'' {\em ApJ} {\bf 593}, p.~L147, 2003.

\bibitem{ResBoostApJ2003}
D.~{Erskine}, J.~{Edelstein}, M.~{Feuerstein}, and B.~{Welsh}, ``{High
  Resolution Broadband Spectroscopy using an Externally Dispersed
  Interferometer},'' {\em ApJ} {\bf 592}, pp.~L103--L106, 2003.

\bibitem{SPIEscot}
D.~J. {Erskine} and J.~{Edelstein}, ``{Interferometric Resolution Boosting for
  Spectrographs},'' in {\em Ground-based Instrumentation for Astronomy. Edited
  by Alan F. M. Moorwood and Iye Masanori. Proceedings of the SPIE, Volume
  5492, pp. 190-199 (2004).},  pp.~190--199, Sept. 2004.

\bibitem{GeAAS2005}
J.~{Ge}, J.~{van Eyken}, S.~{Mahadevan}, C.~{DeWitt}, R.~{Cohen}, A.~{Vanden
  Heuvel}, S.~{Fleming}, P.~{Guo}, S.~{Kane}, G.~{Henry}, G.~{Israelian}, and
  E.~{Martin}, ``{The First Extrasolar Planet Discovered with A New Generation
  High Throughput Doppler Instrument},'' {\em American Astronomical Society
  Meeting Abstracts} {\bf 207}, pp.~--+, 2005.

\end{thebibliography}
\bibliographystyle{spiebib}   
 
 \end{document}